\documentclass{article}

\usepackage{arxiv}
\usepackage[utf8]{inputenc} 
\usepackage[T1]{fontenc}    
\usepackage{hyperref}       
\usepackage{url}            
\usepackage{booktabs}       
\usepackage{amsfonts}       
\usepackage{nicefrac}       
\usepackage{microtype}      
\usepackage{lipsum}		
\usepackage{natbib}
\usepackage{doi}
\usepackage{amsmath,amssymb,amsthm,amsfonts}
\usepackage{algorithmic,algorithm, array}
\usepackage{textcomp,stfloats,verbatim}
\usepackage{graphicx,xcolor,tikz}
\usepackage{authblk}
\usetikzlibrary{patterns}

\newtheorem{proposition}{Proposition}[section]
\newtheorem{remark}{Remark}[section]

\newcommand{\winlength}{\theta} 
\newcommand{\standardwinlength}{L} 
\newcommand{\support}{N} 

\newcommand{\complex}{j}

\newcommand{\idx}{t}
\newcommand{\col}{i} 
\newcommand{\f}{\xi} 
\newcommand{\tap}{\omega}

\newcommand{\spec}{\mathcal{S}}

\definecolor{darkred}{rgb}{.82,0,0}
\definecolor{BleuBleu}{RGB}{0 50 150}
\definecolor{darkgreen}{rgb}{0,.4,0}
\definecolor{lmcolor}{rgb}{0,.6,0}
\definecolor{ymcolor}{rgb}{1,.4,0}
\definecolor{abcolor}{rgb}{0,.6,.9}
\definecolor{mlcolor}{rgb}{0.5,.2,.5}

\title{Differentiable short-time Fourier transform with respect to the hop length}

\date{} 					

\author{ Maxime Leiber{$^{* \dagger}$}, Yosra Marnissi{$^\dagger$}, Axel Barrau{$^\ddagger$}, Mohammed El Badaoui{$^\dagger \mathsection$} \vspace{.5cm} \\
 	{$^*$} INRIA, DI/ENS, PSL Research University \\
 	{$^\dagger$} Safran Tech, Digital Sciences \& Technologies \\
 	{$^\ddagger$} Offroad \\
        {$^\mathsection$} Univ Lyon, UJM-St-Etienne, LASPI \\
}

\renewcommand{\headeright}{}
\renewcommand{\undertitle}{Accepted for IEEE SSP workshop 2023}
\renewcommand{\shorttitle}{Differentiable short-time Fourier transform}

\hypersetup{
pdftitle={Differentiable short-time Fourier transform},
pdfsubject={q-bio.NC, q-bio.QM},
pdfauthor={David S.~Hippocampus, Elias D.~Striatum},
pdfkeywords={time-frequency representation, short-time Fourier transform, spectrogram, differentiable, adaptive, window length, hop length, gradient descent},
}

\begin{document}
\maketitle

\begin{abstract}
In this paper, we propose a differentiable version of the short-time Fourier transform (STFT) that allows for gradient-based optimization of the hop length or the frame temporal position by making these parameters continuous. Our approach provides improved control over the temporal positioning of frames, as the continuous nature of the hop length allows for a more finely-tuned optimization. Furthermore, our contribution enables the use of optimization methods such as gradient descent, which are more computationally efficient than conventional discrete optimization methods. Our differentiable STFT can also be easily integrated into existing algorithms and neural networks. We present a simulated illustration to demonstrate the efficacy of our approach and to garner interest from the research community.
\end{abstract}


\section{Introduction}
\label{sec:1_}
The short-time Fourier transform (STFT) is a frequently used tool for analyzing non-stationary digital signals in various fields including audio \cite{stafford1998long}, medicine \cite{huang2019ecg}, and vibration analysis \cite{leclere2016multi}. Spectrograms, which are obtained from the STFT magnitude, are essential for visualizing, understanding, and processing non-stationary signals in time-frequency representation. 

The STFT parameters, including tapering function, window length, and hop length, are critical and dependent on the application and signal characteristics. The tapering function balances frequency resolution and spectral leakage, with a narrower main lobe providing better frequency resolution at the expense of increased spectral leakage, and a wider main lobe reducing spectral leakage but decreasing frequency resolution. The Hann or Hamming window is a common starting point, but the best choice depends on the application's specific requirements. Actually, most studies on STFT parameters have focused on the choice of the window length, as it determines the time-frequency resolution trade-off. A shorter window length provides better time resolution but poor frequency resolution. Conversely, a longer window length provides better frequency resolution but poor time resolution. To provide more precise control over temporal and frequency resolution based on the local characteristics of the input signal, researchers have proposed using variable-length windows. These methods are known as S transform \cite{Stockwell,Sejdic,Moukadem} and adaptive STFTs \cite{astft,kwok2000improved,Zhong,Pei,Zhu}. Often, the optimal window lengths in the 2D plane are chosen to favor a particular criterion of sparsity \cite{Zhao, Intelligent, Pei, Meignen}. Since these parameters are discrete, finding the optimal tuning is usually done using grid search or trial-and-error, which can be time-consuming. Differentiable versions of the STFT have been recently proposed \cite{Zhao, dstft, gretsi, dastft} making the window length a continuous parameter that can be optimized by gradient descent. 

The third parameter in STFT, the hop length, controls the trade-off between temporal resolution and computational cost. The smaller the hop length is, the higher the temporal resolution is, allowing for more detailed changes in frequency content over time. Conversely, a larger hop length provides lower temporal resolution, which may be sufficient in applications such as music classification \cite{choi2017convolutional} and sound event detection \cite{parascandolo2016recurrent} where the focus is on identifying the presence of certain sound events rather than their exact temporal location. Additionally, in machine learning applications, the hop length choice can significantly impact model performance as a shorter hop length provides more detailed information about data but increases computational complexity. 
Note that, accurately identifying and localizing transient events is critical in many applications, such as bird call classification \cite{acevedo2009automated,koh2019bird} and vibration health monitoring tasks like shock characterization and angular position detection of harmonic sources \cite{sadler1998optimal, chandra2016fault}. Capturing the fine temporal structure of these events provides valuable insights and enables reliable analysis and diagnosis.
However, accurate localization of these events can be challenging if the analysis window is not aligned with the beginning of the target components, leading to energy leakage in the spectrogram. Therefore, selecting the appropriate hop length is critical to capture the temporal dynamics of the signal, such as the attack and decay of the target components.

In summary, the optimal hop length value in spectrograms depends on the specific application and input signal characteristics. To the best of our knowledge, there has not been sufficient research on tuning or adapting the hop length to signal characteristics, as most applications use a fixed hop length. These fixed hop lengths are often set to default values in commonly used signal processing libraries or selected empirically by trial-and-error. 

Our main contribution in this paper is the development of a differentiable version of the STFT, which allows for the easy optimization of the hop length (or frame index) per-frame using gradient descent. This is achieved by modifying the definition of the STFT operator to make the hop length a continuous parameter, with respect to which the STFT values can be differentiated. We utilize the continuity of the tapering function to differentiate with respect to the frame temporal position. The STFT operation is mathematically differentiable, and we provide the necessary calculations and formulas for propagation and backpropagation.

This paper is organized as follows: in Section \ref{sec:2_}, we provide definitions and notations for the STFT and the differentiable STFT. In Section \ref{sec:3_}, we introduce our modified differentiable STFT with respect to the hop length. In Section \ref{sec:4_}, we demonstrate the effectiveness of our approach with a simulated illustration. Finally in Section \ref{sec:5_}, we conclude with final remarks.

\section{Background}
\label{sec:2_}
\subsection{Short-Time Fourier Transform}

Throughout this paper, we refer to the short-time Fourier transform (STFT) as the operation that takes a one-dimensional signal $s[t]$ as input and returns a one-dimensional matrix $\spec[\col,\f]$. Each column $\spec[\col,:]$ of the STFT is the discrete Fourier transform (DFT) of a slice of length $\standardwinlength$ of the signal $s$,  starting from an index $b_\col$ and ending at an index $b_\col + \standardwinlength-1$, multiplied by a tapering function $\tap_\standardwinlength$ of length $\standardwinlength$. The STFT can be mathematically written as:
\begin{equation}
\label{eq::eq1}
\begin{aligned}
\spec[\col,\f] = \sum_{k=0}^{\standardwinlength-1} \tap_\standardwinlength[k]s\left[b_\col+k\right] e^{\frac{-2\complex\pi k\f}{\standardwinlength}}
 \end{aligned}
\end{equation}
The starting indices $b_i$ of the time intervals on which spectra are computed are usually equally spaced, so we only need to set the first index $b_0$ and the spacing $\Delta b$ between $b_i$ and $b_{i+1}$. There are several choices of tapering function $\tap_\standardwinlength$ that can be used, such as the Gaussian and Hann windows.

\subsection{Differentiable Short-Time Fourier Transform}

Differentiable short-time Fourier transform (DSTFT) has been proposed in previous works \cite{dstft, gretsi, dastft}. It modifies the definition of the STFT operator by making the window length a continuous parameter, enabling spectrogram values to be easily differentiated with respect to this parameter  with gradient descent. This differentiable STFT can be readily integrated into any neural networks involving spectrograms, where the STFT can be made into a layer whose weights are the window lengths of the STFT, or can be optimized on-the-fly based on a criterion. 
The idea behind DSTFT is to separate the window length $\standardwinlength$ into an \emph{integer numerical window support} $\support$ and a \emph{continuous time resolution} $\winlength$. Moreover, it is possible to define a continuous window length for each bin of the STFT, which provides a temporal resolution $\winlength_{\col,\f}$ for each frame $\col$ and frequency $\f$:

\begin{equation}
\label{eq::eq2}
\spec[\col,\f] = \sum_{k=-\infty}^{+\infty} \tap_{\support,\winlength_{\col \f}}[k] s\left[\idx_\col +k \right] e^{\frac{-2\complex\pi k\f}{\support}},
\end{equation}
where $\tap_{\support,\winlength_{\col \f}}$ is a tapering function defined on $[0, N-1]$ (is zeros outside this interval) but taking non-zero (or non-negligible) values in the interval $\left[\frac{\support-1-\winlength_{\col \f}}{2} , \frac{\support-1+\winlength_{\col \f}}{2}\right]$.

\section{Proposed method}
\label{sec:3_}


In the same spirit as the differentiable STFT with respect to the window length, this section proposes a differentiable version of the STFT with respect to the hop length, allowing for optimization with gradient descent either locally on-the-fly (online) by minimizing an adaptive criteria for each signal, or globally (offline) by minimizing an overall performance measure of a given task over a dataset.

In the following, we assume that, the tapering functions $\tap_{\support,\winlength_{\col \f}}$ are continuous and differentiable, with continuous derivatives with respect to time $t$ and window length $\winlength$. Examples of such functions are the Hann (Eq. \ref{eq::eq3}) 
and the Gaussian (Eq. \ref{eq::eq4}) functions, which are defined in the support $[0, N-1]$ and have been presented in previous works on the DSTFT \cite{dstft, gretsi, dastft, Zhao}:

\begin{align}
\label{eq::eq3}
&\tap_{\support, \winlength}[k] = \frac{1}{2} - \frac{1}{2} \cos \left( \frac{2 \pi k}{\winlength} \right) 
\end{align}
\begin{align}
\label{eq::eq4}
&\tap_{\support, \winlength}[k] = \exp \left( - \frac{k^2}{ 2 (\frac{\winlength}{6})^2 } \right)
\end{align}

\begin{figure}
\centering
\begin{tikzpicture}[scale=1.5]
\draw[help lines, very thin, color=gray!50, dashed] (.1,-1.6) grid (7.1,3.8);
\draw[domain=.4:7.1, smooth] plot (\x,{3.3+ .2*sin( 10 * \x r)}); 
\draw (.05,3.25) node[below, scale=1.5] {$s$};
\newcommand*{\rotatecurvearrowleft}{\mathbin{\rotatebox{90}{$\curvearrowleft$}}}
\draw (.45,2.85) node[scale=2] {$ \rotatecurvearrowleft $}; 
\draw (.05,2.8) node[scale=1.2] {$\times$}; 

\draw[domain=.9:3.1, color=BleuBleu, ] plot (\x, { 2 +  .5*(1-cos(2*pi*(\x-.9)/2.2 r))  });
\draw[domain=0:.9, color=BleuBleu] plot (\x, 2);
\draw[domain=3.1:4, color=BleuBleu] plot (\x, 2);
\draw[domain=1.6:3.8, color=BleuBleu, dash pattern=on 5pt off 5pt] plot (\x, { 2+  .5*(1-cos(2*pi*(\x-.7-.9)/2.2 r))  });
\draw[domain=0.:1.6, color=BleuBleu, dash pattern=on 5pt off 5pt] plot (\x, 2);
\draw[domain=3.8:4., color=BleuBleu, dash pattern=on 5pt off 5pt] plot (\x, 2);
\draw [stealth-stealth, darkgreen](0,1.75) -- (4,1.75);
\draw (2,1.75) node[below, darkgreen] {$\support$}; 

\draw[domain=4.5:5.9, color=BleuBleu!80 ] plot (\x, {1.25 +   .5*(1-cos(2*pi*(\x-4.5)/1.4 r))  });
\draw[domain=3.:4.5, color=BleuBleu!80] plot (\x, 1.25);
\draw[domain=5.9:7., color=BleuBleu!80] plot (\x, 1.25);
\draw[domain=4.2:5.6, color=BleuBleu!80 , dash pattern=on 3pt off 3pt] plot (\x, {1.25+  .5*(1-cos(2*pi*(\x-4.2)/1.4 r))  });
\draw[domain=2.:4.2, color=BleuBleu!80 , dash pattern=on 3pt off 3pt] plot (\x, 1.25);
\draw[domain=5.6:6, color=BleuBleu!80, dash pattern=on 3pt off 3pt] plot (\x, 1.25);
\draw [stealth-stealth, darkgreen](3,1) -- (7,1);
\draw [stealth-stealth, darkgreen, dash pattern=on 5pt off 5pt](2, 0.9) -- (6, 0.9);
\draw (4.5, .9) node[below, darkgreen] {$\support$}; 

\draw [dashed, gray](0, -1.2) -- (0, 1.9);
\draw [dashed, gray](0.7, -1.2) -- (0.7, 1.9);
\draw [dashed, gray](3.2, -1.2) -- (3.2, 1.2);
\draw [dashed, gray](2.9, -1.2) -- (2.9, 1.2);

\draw [darkred, -stealth](0, 1.2) -- (0.7, 1.2);
\draw (0,1.2) node[left, darkred] {$\idx_\col^{(j)}$};
\draw (0.7,1.2) node[right, darkred] {$\idx_\col^{(j+1)}$};

\draw [darkred, stealth-](2.9, .5) -- (3.2, .5);
\draw (3.2, .5) node[right, darkred] {$\idx_{\col+1}^{(j)}$};
\draw (2.9, .5) node[left, darkred] {$\idx_{\col+1}^{(j+1)}$};

\draw [stealth-stealth, color=mlcolor](0, 0) -- (3.2, 0);
\draw (1.6, 0) node[above, color=mlcolor] {  $H_{i+1}^{(j)}$}; 
\draw [stealth-stealth, color=mlcolor](0.7, -.2) -- (2.9, -.2);
\draw (1.8, -.2) node[below, color=mlcolor] {$H_{i+1}^{(j+1)}$}; 

\draw [dashed, gray](0.9, -1.2) -- (0.9, 1.9);
\draw [dashed, gray](1.6, -1.2) -- (1.6, 1.9);
\draw [dashed, gray](4.5, -1.2) -- (4.5, 1.2);
\draw [dashed, gray](4.2, -1.2) -- (4.2, 1.2);

\draw [stealth-stealth](0.9, -1.) -- (4.5, -1.);
\draw (2.7, -1.) node[above] {$\tilde{H}_{i+1}^{(j)}$}; 
\draw [stealth-stealth](1.6, -1.2) -- (4.2, -1.2);
\draw (2.9, -1.2) node[below] { $\tilde{H}_{i+1}^{(j+1)}$}; 

\draw (1.1, 2.5) node[above, color=BleuBleu ] { $\tap_{\support,\winlength_{\col \f}}^{(j)}$ };
\draw (3.7, 2.5) node[above, color=BleuBleu ] { $\tap_{\support,\winlength_{\col \f}}^{(j+1)}$ };
\draw (6, 2.) node[above, color=BleuBleu ] { $\tap_{\support,\winlength_{\col+1 \f}}^{(j)}$ };



\end{tikzpicture}
\caption{Differentiable STFT: The position of the tapering windows can smoothly shift along the time axis, while the window support starts at the integer part of the temporal position of the tapering windows. The exponent $j$ denotes the iteration in the gradient descent optimizer.}
\label{fig:fig1}
\end{figure}
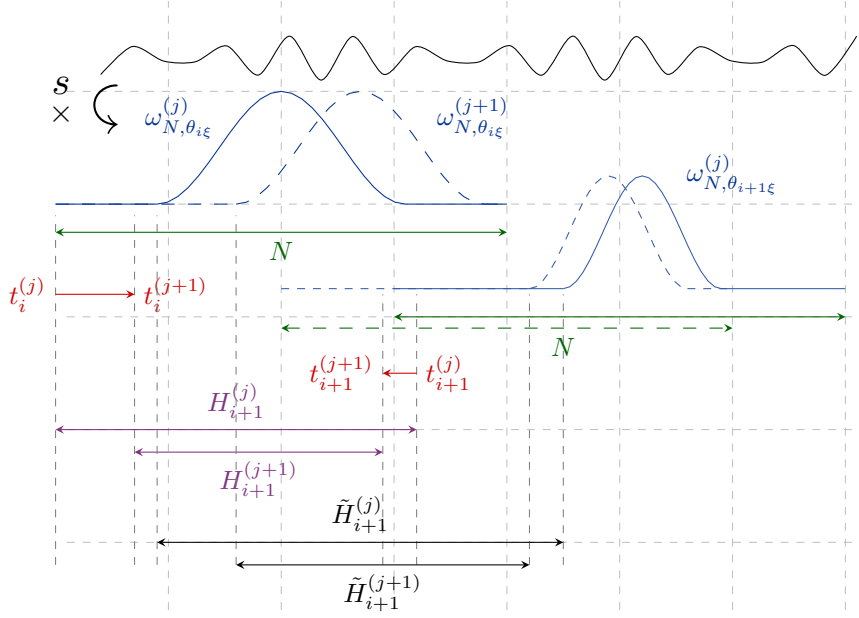
The STFT can be differentiated with respect to the hop length $H_i$ by equivalently differentiating the STFT spectrum $\spec[\col,\f]$ with respect to the temporal position $\idx_{\col}$, where the hop length is the difference between two consecutive frames i.e $H_i = t_{i}-t_{i-1}$, $H_1 = t_1$. This is possible because the tapering function $\tap_{\support,\winlength_{\col,\f}}$ is continuous and differentiable with respect to the temporal position $\idx_{\col}$. However, there is one constraint: the analysis window must start on integer values since the signal is discrete. To overcome this constraint, the window is started at $\lfloor \idx_{\col} \rfloor$ as displayed in Fig. \ref{fig:fig1}, and the effect of the integer part is compensated by a small shift $ \{ \idx_{\col} \} = \idx_{\col} - \lfloor \idx_{\col} \rfloor$ in the argument of $\tap_{\support,\winlength_{\col,\f}}$ and a factor $e^{2\complex\pi \{ \idx_{\col} \} \f / \support}$. This makes this modified version of the STFT differentiable with respect to the frame index $\idx_\col$ (and equivalently the hop length $H_i$) and the window length $\winlength_{\col,\f}$:
\begin{equation}
\label{eq::eq7}
\spec[\col,\f] = \sum_{k=\infty}^{+\infty} \tap_{\support,\winlength_{\col \f}}[k - \{ \idx_{\col} \}] s\left[\lfloor \idx_{\col} \rfloor +k \right] e^{\frac{-2\complex\pi (k- \{ \idx_{\col} \} ) \f}{\support}}
\end{equation}
The rest of this section will be devoted to the justification of the formula presented in Eq. (\ref{eq::eq7}) by showing that it is continuous and differentiable with respect to $\idx_\col$.



\begin{proposition}
The STFT defined by Eq. (\ref{eq::eq7}) is everywhere continuous with respect to the frame index $\idx_\col$.
\end{proposition}
\begin{proof}
The integer and fractional part functions are continuous everywhere except at integers.  Therefore, Eq. (\ref{eq::eq7}) is continuous everywhere except possibly when $\idx_\col$ is an integer value $p \in \mathbb{N}$. We will prove that it is also continuous in this case by looking at its left and right limits, i.e. for $\idx_\col = p \pm \epsilon$ as $\epsilon \rightarrow 0$.

For $\idx_\col = p - \epsilon$, we have $\lfloor \idx_{\col} \rfloor = p-1$ and $ \{ \idx_{\col} \} = 1 - \epsilon$, which implies:
\begin{align}
\label{eq::eq8}
\spec[i,\f] &= \sum_{k=-\infty}^{+\infty} \tap_{\support,\winlength_{\col \f}} [k-1+\epsilon] s[p-1+k] e^{\frac{-2\complex\pi (k-1+\epsilon)f}{\support}} \nonumber\\ 
 &= \sum_{k=-\infty}^{+\infty} \tap_{\support,\winlength_{\col \f}}[k+ \epsilon] s[p+k] e^{\frac{-2\complex\pi (k+ \epsilon)f}{\support}}
\end{align}


For $\idx_\col = p + \epsilon$, we have $\lfloor \idx_{\col} \rfloor = p$ and $ \{ \idx_\col \} = \epsilon$, which implies:
\begin{align}
\label{eq::eq9}
\spec[i,\f] =\sum_{k=-\infty}^{+\infty} \tap_{\support,\winlength_{\col \f}}[k- \epsilon] s[p+k] e^{\frac{-2\complex\pi (k- \epsilon)f}{\support}}
\end{align}

So the limit is the same for $\idx_\col = p \pm \epsilon$ when $\epsilon \rightarrow 0$.
\end{proof}

We have proved that our proposed STFT is continuous with respect to the frame index  $\idx_\col$. Let us now consider its differentiability with respect to $\idx_\col$. Indeed, the integer and fractional part functions are differentiable everywhere except on integers. So Eq. (\ref{eq::eq7}) is differentiable everywhere except possibly when $\idx_\col$ takes an integer value. Using the fact that $\frac{\partial\lfloor \idx_{\col} \rfloor}{ \idx_{\col}}=0$ and $\frac{\partial\{ \idx_{\col} \} }{ \idx_{\col}}=1$ for non integers $\idx_{\col}$, we can easily show that: 
\begin{align}
\label{eq::eq10}
\frac{\partial \spec(\col,\f)}{\partial  \idx_\col} = \sum_{k=-\infty}^{ +\infty} \tilde \tap_{\support,\winlength_{\col \f}}[k- \{ \idx_\col \}] s[\lfloor \idx_{\col} \rfloor +k] e^{  \frac{-2\complex \pi (k -  \{ \idx_{\col} \} ) \f}{ \support}}
\end{align}
with $\tilde  \tap_{\support,\winlength_{\col \f}} = - \dot{\tap}_{\support,\winlength_{\col \f}}+  2\complex \pi  \f \tap_{\support,\winlength_{\col \f}}$ and $\dot{\tap}_{\support,\winlength_{\col \f}}(u) = \frac{\partial \tap_{\support,\winlength_{\col \f}}(u)}{\partial u} $. 
Extension to any value of $\idx_{\col}$ is the point of Prop. \ref{prop2} below.

\begin{proposition}
\label{prop2}
The modified STFT defined by Eq. (\ref{eq::eq7}) is everywhere differentiable w.r.t. $\idx_\col$.
\end{proposition}
\begin{proof}


Since the window $\tap_{\support,\winlength_{\col \f}}$ is continuous function with continuous derivatives with respect to time variable, $\tilde  \tap_{\support,\winlength_{\col \f}}$ is still a continuous function with continuous derivatives with respect to time variable\footnote{Both real and imaginary parts are continuous and differentiable with continuous derivatives with respect to time variable.}. Eq. (\ref{eq::eq10}) has then the same shape as Eq. (\ref{eq::eq7}) meaning a similar reasoning will show that the derivative Eq. (\ref{eq::eq10}) converges to the same limits when $ \theta$ goes to an integer by lower or upper values. As a consequence, the proposed STFT is differentiable everywhere.




\end{proof}
\begin{remark}
Eq. \ref{eq::eq10} defines the derivatives with respect to the frame index $\idx_\col$. As the hop length $H_\col$ is just the difference between consecutive frames indexes $\idx_\col$ and $\idx_{\col-1}$, the derivatives with respect to hop length is straightforward. 
\end{remark}

\begin{remark}
Since the window length $\winlength_{\col, \f}$ can vary within the support interval $[0, \support]$, it would be more appropriate to define a "true" hop length $\tilde{H}_{\col, \f}$ that is computed as the difference between the positive starting points of two consecutive windows.  This can be calculated using the following formula (see Fig. \ref{fig:fig1}):
\begin{equation}
\label{eq::eq11}
\begin{aligned}
&\tilde{H}_{1, \f} = H_1 + \frac{ \support - \winlength_{1, \f} }{2} \\
&\tilde{H}_{\col, \f} = H_\col + \frac{\winlength_{\col, \f} - \winlength_{\col-1, \f}}{2}  \quad \forall \col >1 
 \end{aligned}
\end{equation}
Then, the expression of derivatives with respect to $\tilde{H}_{\col, \f}$ is straightforward. One compelling reason to use the "true" hop length is that the window length may vary between frames in adaptive STFT with a time-varying window length. 
\end{remark}

\begin{remark}
In our differentiable STFT definition, we use one hop length parameter per frame. It is however possible to share parameters and suppose time invariance as in the classic STFT case to share only one window length and hop length.
\end{remark}

\section{Applications}
\label{sec:4_}
\begin{figure*}[t!]
    \centering
    \begin{tikzpicture}
        \node[inner sep=0pt] (img1) {\includegraphics[width=8.5cm]{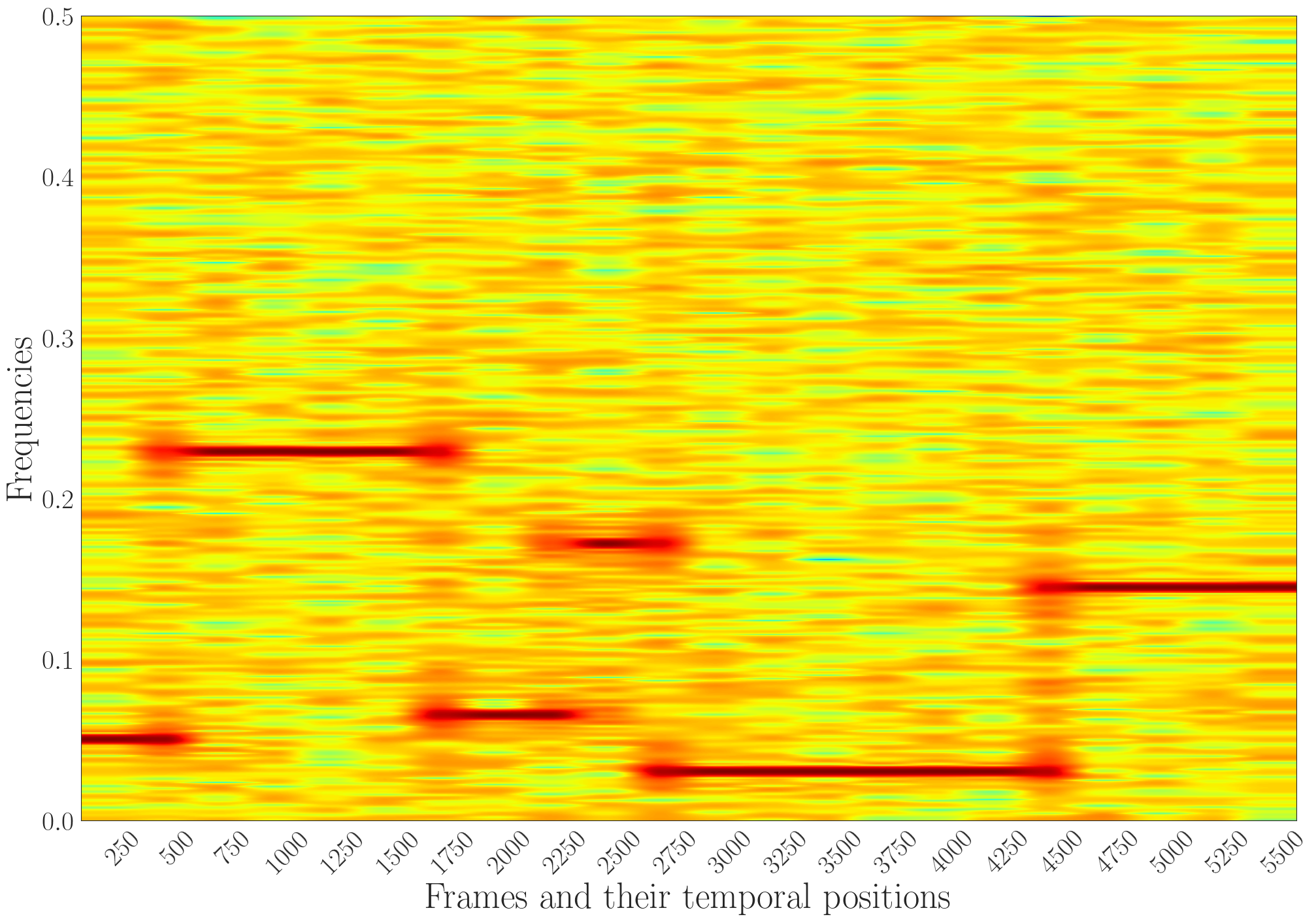}};
        \node[inner sep=0pt] (img2) at (img1.east) [right=.5cm] {\includegraphics[width=8.5cm]{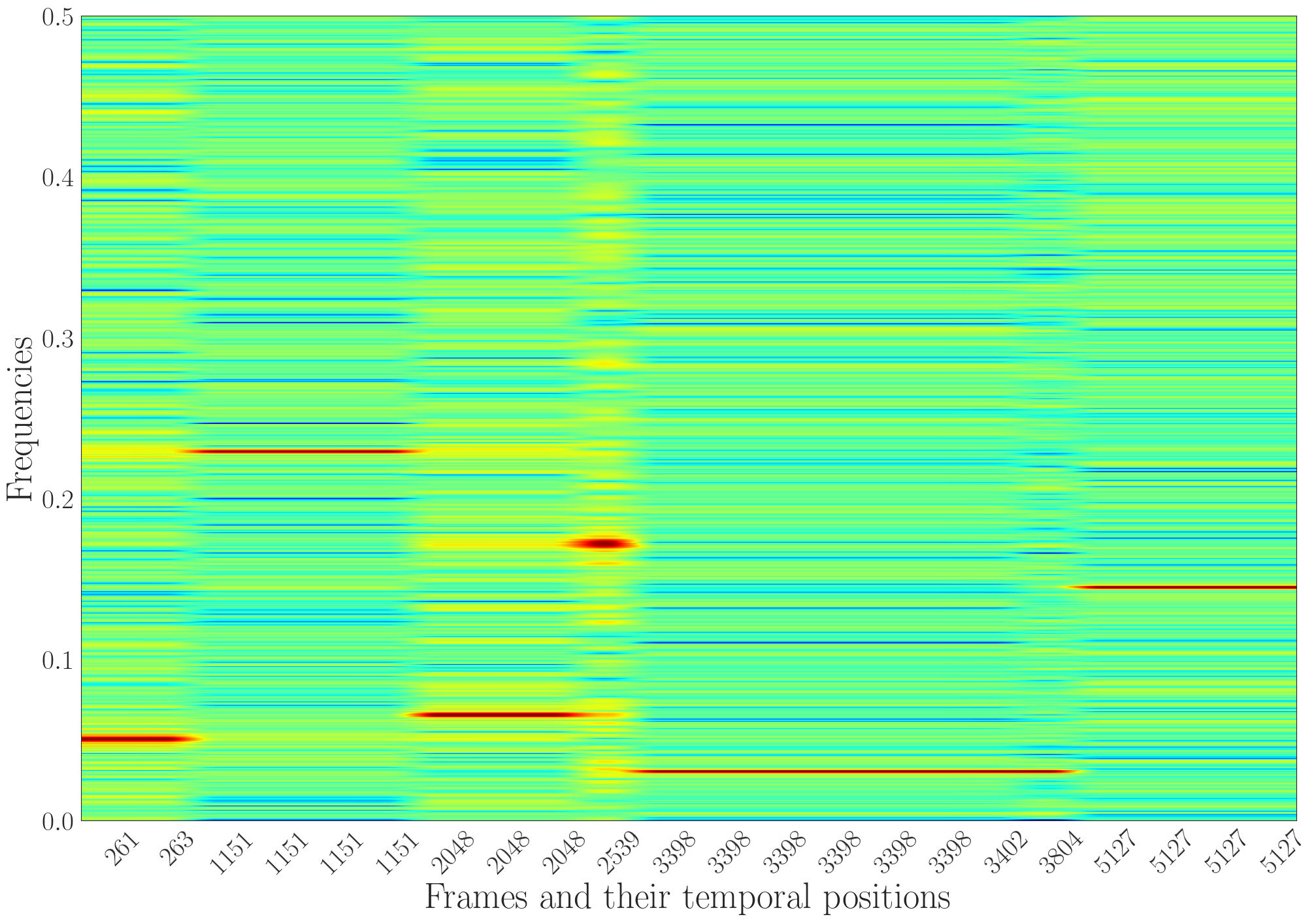}};
        \node[inner sep=0pt] (img3) at (img1.south) [below=0.0cm] {\includegraphics[width=8.5cm]{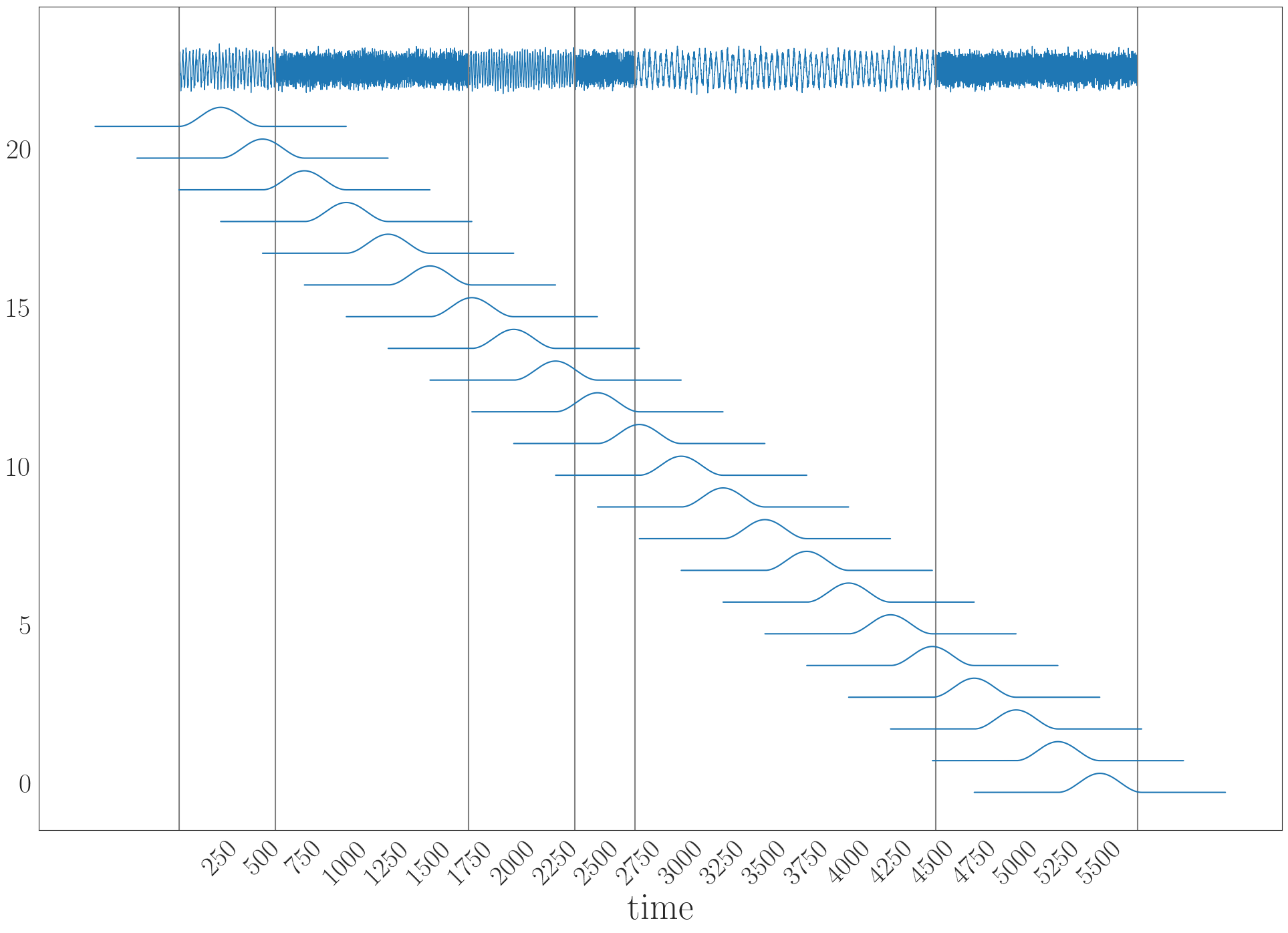}};
        \node[inner sep=0pt] (img4) at (img3.east) [right=.5cm] {\includegraphics[width=8.5cm]{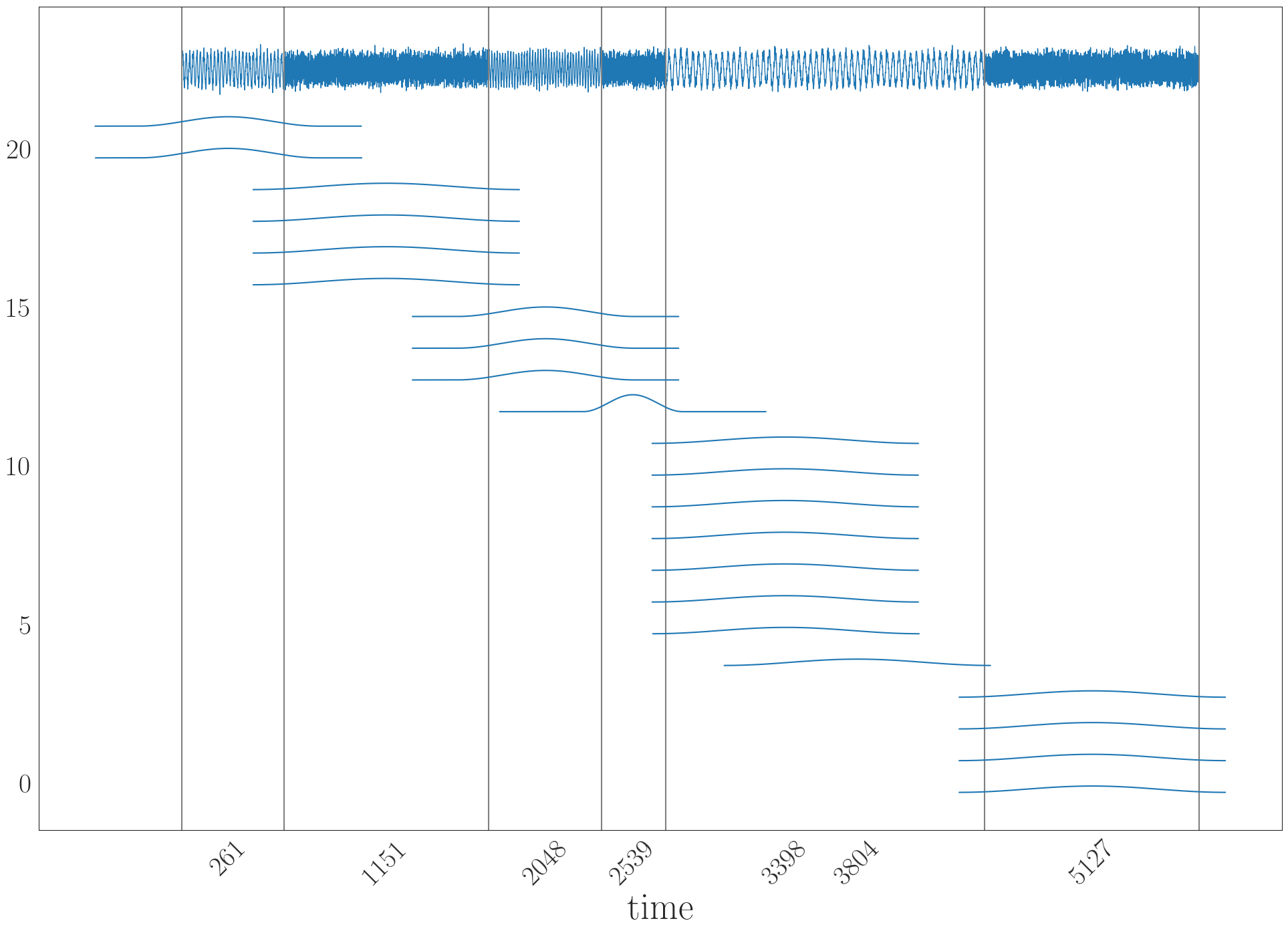}};
    \end{tikzpicture}
    \caption{On the left side, we have the classical STFT and its frame temporal position along the signal. On the right side, we have the DSTFT with time-varying frame temporal position and window length, and the DSTFT frame temporal position along the signal.}
    \label{fig:fig2}
\end{figure*}


In this section, we will show on a simulated signal that differentiable STFT with respect to both window length and hop length (or frame temporal position) can be of immediate interest and deserves more attention. We simulate a sinus of different frequencies with different temporal length. Classical STFT uses frames that are uniformly spread along the signal, which may not be the optimal positioning to localize frequency changes as frequencies have a variable length. 
We initialize our differentiable STFT as classical STFT. In this application, we consider using time-varying window length. We then optimize the windows' temporal position and lengths by gradient descent according to a given criterion. More specifically, our goal is to find windows that are well-distributed over the overall signal, ensuring that there are no gaps and loss of information, and that allow for minimal energy leakage while achieving a better concentration of energy in the time-frequency plane. To promote energy concentration in each frame, we maximize the kurtosis of frame spectrum: 
\vspace{-.2cm}
\begin{equation}
   \mathcal{K}(\winlength_{\col}, \tilde{H}_{\col}) = \frac{1}{\sum_{\col=1}^{T} \omega_{\col}}\sum_{\col=1}^{T} \omega_{\col} \frac{\mathbb{E}_\f[\spec[\col, \f]^4]}{\mathbb{E}_\f[\spec[\col, \f]^2]^2}
\end{equation}
where 
$\omega_{\col}$ is a weight used to minimize the contribution of windows that share the same segment of the signal. More specifically, we set $\omega_{1}=\idx_{2} - \idx_{1}$, $\omega_{T}=\idx_{T} - \idx_{T-1}$ and $\omega_{\col} = \frac{\idx_{\col+1} - \idx_{\col-1}}{2}$ for every $\col \in \{2, \ldots, T-1 \}$. 
To avoid information loss, we also want to maximize the spectrogram coverage \footnote{Let's recall that in cases where the window's support extends beyond the signal, we zero-pad the signal for the calculation of the STFT. However, it's important to note that the coverage metric only takes into account the part of the window that overlaps with the signal.}: 
\vspace{-.2cm}
\begin{equation}
    \mathcal{C}(\winlength_{\col}, \tilde{H}_{\col})= \dfrac{\sum_{\col=1}^{T} \min\left(\winlength_\col, \tilde{H}_{\col +1}\right)}{M}
\end{equation}
with $\tilde{H}_{T+1}=+\infty$ and $M$ is the signal length. Finally, we use a multi-objective optimization to maximize both the energy concentration $\mathcal{K}(\winlength_{\col}, \tilde{H}_{\col})$ and the coverage of the spectrogram $\mathcal{C}(\winlength_{\col}, \tilde{H}_{\col})$\cite{desideri2012multiple, sener2018multi}. \footnote{We have shared the notebook to reproduce the experiment at \url{https://github.com/maxime-leiber/dstft}.}


We observe that the spectrogram obtained at the end of the optimization has a better concentration of energy due to reduced spectral leakage resulting from the proper positioning of frames, which are no longer straddling two different frequencies, as shown in Fig. \ref{fig:fig2}. However, caution must be exercised when reading the time-axis of the resulting time-frequency representation as the distribution of frames is no longer uniform.

\section{Conclusion}
\label{sec:5_}
In conclusion, we have introduced a novel modification of the DSTFT, which makes this operation differentiable with respect to the hop length or frame temporal position. Through an example, we have demonstrated the benefit of using our differentiable STFT. Our proposed approach provides improved control over the temporal positioning of frames and enables the use of more computationally efficient optimization methods, such as gradient descent. Our differentiable STFT can also be easily integrated into existing algorithms and neural networks.

\clearpage

\bibliographystyle{unsrtnat}
\bibliography{refs}  

\begin{thebibliography}{25}
\providecommand{\natexlab}[1]{#1}
\providecommand{\url}[1]{\texttt{#1}}
\expandafter\ifx\csname urlstyle\endcsname\relax
  \providecommand{\doi}[1]{doi: #1}\else
  \providecommand{\doi}{doi: \begingroup \urlstyle{rm}\Url}\fi

\bibitem[Stafford et~al.(1998)Stafford, Fox, and Clark]{stafford1998long}
Kathleen~M Stafford, Christopher~G Fox, and David~S Clark.
\newblock Long-range acoustic detection and localization of blue whale calls in
  the northeast pacific ocean.
\newblock \emph{The Journal of the Acoustical Society of America}, 104\penalty0
  (6):\penalty0 3616--3625, 1998.

\bibitem[Huang et~al.(2019)Huang, Chen, Yao, and He]{huang2019ecg}
Jingshan Huang, Binqiang Chen, Bin Yao, and Wangpeng He.
\newblock Ecg arrhythmia classification using stft-based spectrogram and
  convolutional neural network.
\newblock \emph{IEEE access}, 7:\penalty0 92871--92880, 2019.

\bibitem[Lecl{\`e}re et~al.(2016)Lecl{\`e}re, Andr{\'e}, and
  Antoni]{leclere2016multi}
Q.~Lecl{\`e}re, H.~Andr{\'e}, and J.~Antoni.
\newblock A multi-order probabilistic approach for instantaneous angular speed
  tracking debriefing of the cmmno14 diagnosis contest.
\newblock \emph{Mechanical Systems and Signal Processing}, 81:\penalty0
  375--386, 2016.

\bibitem[Stockwell et~al.(1996)Stockwell, Mansinha, and Lowe]{Stockwell}
R.~Stockwell, L.~Mansinha, and R.~Lowe.
\newblock Localization of the complex spectrum: the {S} transform.
\newblock \emph{IEEE transactions on Signal Processing}, 44\penalty0
  (4):\penalty0 998--1001, 1996.

\bibitem[Sejdi{\'c} et~al.(2007)Sejdi{\'c}, Djurovi{\'c}, and Jiang]{Sejdic}
Ervin Sejdi{\'c}, Igor Djurovi{\'c}, and Jin Jiang.
\newblock A window width optimized s-transform.
\newblock \emph{EURASIP Journal on Advances in Signal Processing},
  2008:\penalty0 1--13, 2007.

\bibitem[Moukadem et~al.(2015)Moukadem, Bouguila, Abdeslam, and
  Dieterlen]{Moukadem}
Ali Moukadem, Zied Bouguila, Djaffar~Ould Abdeslam, and Alain Dieterlen.
\newblock A new optimized stockwell transform applied on synthetic and real
  non-stationary signals.
\newblock \emph{Digital Signal Processing}, 46:\penalty0 226--238, 2015.

\bibitem[Czerwinski and Jones(1997)]{astft}
R.~N. Czerwinski and D.~L. Jones.
\newblock Adaptive short-time fourier analysis.
\newblock In \emph{IEEE Signal Processing Letters}, volume~4, pages 42--45,
  1997.

\bibitem[Kwok and Jones(2000)]{kwok2000improved}
Henry~K Kwok and Douglas~L Jones.
\newblock Improved instantaneous frequency estimation using an adaptive
  short-time fourier transform.
\newblock \emph{IEEE Transactions on Signal Processing}, 48:\penalty0
  2964--2972, 2000.

\bibitem[Zhong and Huang(2010)]{Zhong}
J.~Zhong and Y.~Huang.
\newblock Time-frequency representation based on an adaptive short-time fourier
  transform.
\newblock In \emph{IEEE Transactions on Signal Processing}, volume~58, pages
  5118--5128, 2010.

\bibitem[Pei and Huang(2012)]{Pei}
S.~Pei and S.~Huang.
\newblock {STFT} with adaptive window width based on the chirp rate.
\newblock \emph{IEEE Transactions on Signal Processing}, 60\penalty0
  (8):\penalty0 4065--4080, 2012.

\bibitem[Zhu et~al.(2015)Zhu, Zhang, and Qi]{Zhu}
Mingzhe Zhu, Xinliang Zhang, and Yue Qi.
\newblock An adaptive stft using energy concentration optimization.
\newblock In \emph{International Conference on Information, Communications and
  Signal Processing (ICICS)}, pages 1--4, 2015.

\bibitem[Zhao et~al.(2021)Zhao, Subramani, and Smaragdis]{Zhao}
A.~Zhao, K.~Subramani, and P.~Smaragdis.
\newblock Optimizing short-time fourier transform parameters via gradient
  descent.
\newblock In \emph{IEEE International Conference on Acoustics, Speech and
  Signal Processing (ICASSP)}, pages 736--740, 2021.

\bibitem[Jablonski and Dziedziech(2022)]{Intelligent}
Adam Jablonski and Kajetan Dziedziech.
\newblock Intelligent spectrogram--a tool for analysis of complex
  non-stationary signals.
\newblock \emph{Mechanical Systems and Signal Processing}, 167:\penalty0
  108554, 2022.

\bibitem[Meignen et~al.(2020)Meignen, Colominas, and Pham]{Meignen}
S.~Meignen, M.~Colominas, and D.~Pham.
\newblock On the use of r{\'e}nyi entropy for optimal window size computation
  in the short-time fourier transform.
\newblock In \emph{IEEE International Conference on Acoustics, Speech and
  Signal Processing (ICASSP)}, pages 5830--5834, 2020.

\bibitem[Leiber et~al.(2022{\natexlab{a}})Leiber, Barrau, Marnissi, and
  Abboud]{dstft}
M.~Leiber, A.~Barrau, Y.~Marnissi, and D.~Abboud.
\newblock A differentiable short-time fourier transform with respect to the
  window length.
\newblock In \emph{European Signal Processing Conference (EUSIPCO)}, pages
  1392--1396, 2022{\natexlab{a}}.

\bibitem[Leiber et~al.(2022{\natexlab{b}})Leiber, Barrau, Marnissi, Abboud, and
  El~Badaoui]{gretsi}
Maxime Leiber, Axel Barrau, Yosra Marnissi, Dany Abboud, and Mohammed
  El~Badaoui.
\newblock Optimisation de la longueur de fen{\^e}tre du spectrogramme au sein
  d’un r{\'e}seau de neurones.
\newblock In \emph{colloque GRETSI}, 2022{\natexlab{b}}.

\bibitem[Leiber et~al.(2023)Leiber, Marnissi, Barrau, and El~Badoui]{dastft}
M.~Leiber, Y.~Marnissi, A.~Barrau, and M.~El~Badoui.
\newblock Differentiable adaptive short-time fourier transform with respect to
  the window length.
\newblock In \emph{IEEE International Conference on Acoustics, Speech and
  Signal Processing (ICASSP)}, pages 1392--1396, 2023.

\bibitem[Choi et~al.(2017)Choi, Fazekas, Sandler, and
  Cho]{choi2017convolutional}
Keunwoo Choi, Gy{\"o}rgy Fazekas, Mark Sandler, and Kyunghyun Cho.
\newblock Convolutional recurrent neural networks for music classification.
\newblock In \emph{2017 IEEE International conference on acoustics, speech and
  signal processing (ICASSP)}, pages 2392--2396. IEEE, 2017.

\bibitem[Parascandolo et~al.(2016)Parascandolo, Huttunen, and
  Virtanen]{parascandolo2016recurrent}
Giambattista Parascandolo, Heikki Huttunen, and Tuomas Virtanen.
\newblock Recurrent neural networks for polyphonic sound event detection in
  real life recordings.
\newblock In \emph{2016 IEEE international conference on acoustics, speech and
  signal processing (ICASSP)}, pages 6440--6444, 2016.

\bibitem[Acevedo et~al.(2009)Acevedo, Corrada-Bravo, Corrada-Bravo,
  Villanueva-Rivera, and Aide]{acevedo2009automated}
Miguel~A Acevedo, Carlos~J Corrada-Bravo, H{\'e}ctor Corrada-Bravo, Luis~J
  Villanueva-Rivera, and T~Mitchell Aide.
\newblock Automated classification of bird and amphibian calls using machine
  learning: A comparison of methods.
\newblock \emph{Ecological Informatics}, 4\penalty0 (4):\penalty0 206--214,
  2009.

\bibitem[Koh et~al.(2019)Koh, Chang, Tai, Huang, Hsieh, and Liu]{koh2019bird}
Chih-Yuan Koh, Jaw-Yuan Chang, Chiang-Lin Tai, Da-Yo Huang, Han-Hsing Hsieh,
  and Yi-Wen Liu.
\newblock Bird sound classification using convolutional neural networks.
\newblock In \emph{CLEF (Working Notes)}, 2019.

\bibitem[Sadler et~al.(1998)Sadler, Pham, and Sadler]{sadler1998optimal}
Brian~M Sadler, Tien Pham, and Laurel~C Sadler.
\newblock Optimal and wavelet-based shock wave detection and estimation.
\newblock \emph{The Journal of the Acoustical Society of America}, 104\penalty0
  (2):\penalty0 955--963, 1998.

\bibitem[Chandra and Sekhar(2016)]{chandra2016fault}
N.~H. Chandra and A.~Sekhar.
\newblock Fault detection in rotor bearing systems using time frequency
  techniques.
\newblock \emph{Mechanical Systems and Signal Processing}, 72:\penalty0
  105--133, 2016.

\bibitem[D{\'e}sid{\'e}ri(2012)]{desideri2012multiple}
Jean-Antoine D{\'e}sid{\'e}ri.
\newblock Multiple-gradient descent algorithm (mgda) for multiobjective
  optimization.
\newblock \emph{Comptes Rendus Mathematique}, 350\penalty0 (5-6):\penalty0
  313--318, 2012.

\bibitem[Sener and Koltun(2018)]{sener2018multi}
Ozan Sener and Vladlen Koltun.
\newblock Multi-task learning as multi-objective optimization.
\newblock \emph{Advances in neural information processing systems}, 31, 2018.

\end{thebibliography}

\end{document}